\documentclass[10pt,aps,prd,twocolumn,superscriptaddress,nofootinbib,longbibliography]{revtex4-2}
\usepackage{graphicx}
\usepackage{tikz}
\usetikzlibrary{arrows.meta,calc,positioning,decorations.pathmorphing}
\usepackage{dcolumn}
\usepackage{bm}
\usepackage{orcidlink}
\usepackage{multirow}
\usepackage[normalem]{ulem}
\usepackage[dvipsnames]{xcolor}
\usepackage{mathrsfs}
\usepackage{mathtools}
\usepackage{enumitem}
\usepackage{tcolorbox}
\tcbuselibrary{breakable, skins}

\usepackage{hyperref}
\hypersetup{
  colorlinks=true,
  linkcolor=blue,
  citecolor=blue,
  urlcolor=blue
}
\usepackage{cleveref} 

\begin{document}

\title{Josephson Effects in Slowly Rotating Spacetimes}

\author{Nurmukhammed Aytimbetov} 
\email{nurmuxammedzaytimbetov@gmail.com}
\affiliation{Namangan State University, Boburshokh 161, 160107, Uzbekistan}

\author{Reggie C. Pantig \orcidlink{0000-0002-3101-8591}} 
\email{rcpantig@mapua.edu.ph}
\affiliation{Physics Department, School of Foundational Studies and Education, Map\'ua University, 658 Muralla St., Intramuros, Manila 1002, Philippines}

\author{Ali~\"Ovg\"un\orcidlink{0000-0002-9889-342X}}
\email{ali.ovgun@emu.edu.tr}
\affiliation{Physics Department, Eastern Mediterranean University, Famagusta, 99628 North Cyprus via Mersin 10, Turkiye}

\author{Bobomurat Ahmedov \orcidlink{0000-0002-1232-610X}} 
\email{ahmedov@astrin.uz}
\affiliation{Institute for Advanced Studies, New Uzbekistan University,
Movarounnahr str. 1, Tashkent 100000, Uzbekistan}
\affiliation{Institute of Theoretical Physics, National University of Uzbekistan, Tashkent 100174, Uzbekistan}
\affiliation{School of Physics, Harbin Institute of Technology, Harbin 150001, People’s Republic of China}

\author{Javlon Rayimbaev \orcidlink{0000-0001-9293-1838}}
\email{javlon@astrin.uz}
\affiliation{Tashkent State Technical University, Tashkent 100095, Uzbekistan}
\affiliation{University of Tashkent for Applied Sciences, Gavhar Str. 1, Tashkent 700127, Uzbekistan}

\begin{abstract}
We investigate Josephson phenomena in a slowly rotating stationary spacetime, emphasizing the distinct roles of gravitational redshift and rotational frame dragging motivated by [10.1007/JHEP02(2026)006]. Using a covariant formulation based on gauge-invariant phase dynamics and conserved currents within a $3+1$ decomposition, we analyze both AC and DC Josephson effects and interferometric configurations. Restricting attention to linear order in the rotation parameter $a$, and working in the Eulerian/ZAMO frame, we show that in the slow-rotation slicing adopted here the lapse function agrees with its static (Schwarzschild-type) form up to $\mathcal{O}(a^2)$, while rotational effects enter through the shift vector. Consequently, redshift effects on Josephson frequencies and DC critical currents remain unchanged relative to the non-rotating case at $\mathcal{O}(a)$. The AC Josephson relation retains its redshifted structure when expressed in terms of proper voltages and reduces to the standard flat-spacetime form when formulated in terms of asymptotic (Killing-time) observables. Likewise, the DC critical current measured at infinity scales with a single power of the lapse function and is unaffected by rotation at linear order in the absence of azimuthal condensate momentum. Rotational effects become relevant only in configurations sensitive to spatial phase transport or to synchronization with respect to the global time coordinate. In particular, RF-driven interferometric setups can acquire Sagnac-type phase offsets associated with frame dragging, whereas the DC fluxoid constraint remains unshifted at linear order in the present approximation. Our results provide a clean separation between lapse-driven redshift effects and shift-driven rotational contributions in Josephson physics and furnish a consistent framework for superconducting circuits in stationary spacetimes.
\end{abstract}
\maketitle

\section{Introduction}\label{sec:I}

The Josephson effect provides one of the most precise links between quantum phase coherence and macroscopic observables, relating the dynamics of a gauge-invariant condensate phase to electromagnetic control parameters. In flat spacetime, the AC Josephson relation establishes a universal frequency-voltage law, while the DC effect yields a characteristic current-phase relation for weak links. These phenomena underpin modern quantum metrology and superconducting circuit technology \cite{Josephson, Ummarino2020}. For standard discussions of Josephson junction dynamics, microwave response, and superconducting interferometry, see Refs. \cite{Likharev:2022pxz,Clarke_2004}. On the other hand, related gravitationally induced topological phase effects have also been investigated in nontrivial background geometries; in particular, the gravitational Aharonov--Bohm phase was recently studied in \cite{Chiao:2023ezj,Ovgun:2025pwy}.

In curved spacetime, however, the notions of time, energy, and voltage become observer-dependent. In a static gravitational field, clocks redshift, and intensive quantities obey the Tolman-Ehrenfest relations, raising the question of how the Josephson relations are reorganized when superconducting elements are placed at different gravitational potentials. This problem has recently been addressed for static backgrounds, demonstrating that gravitational redshift modifies the mapping between local (proper) and asymptotic observables without altering the local Josephson microphysics \cite{Tolman1930, Dewitt1966}. 

Astrophysical compact objects, however, are generically rotating. Neutron stars and black holes carry angular momentum, and their exterior geometry is more accurately described by a stationary, slowly rotating spacetime rather than by a purely static solution. Rotation introduces qualitatively new features: spacetime becomes stationary rather than static, and inertial frames experience frame dragging \cite{Lense:1918zz,Mashhoon:1984fj,Schiff:1960gh,Everitt:2011hp,Visser2007,Kerr1982}. From the perspective of superconducting phase coherence, this raises a natural and important question: does rotation modify Josephson redshift relations, or does it enter through a distinct physical channel?

In this work, we extend the covariant formulation of Josephson physics from static spacetimes to slowly rotating stationary spacetimes, emphasizing the distinct roles of gravitational redshift and rotational effects. Restricting attention to linear order in the rotation parameter $a$, we exploit the $3 + 1$ decomposition of the metric to disentangle the roles of gravitational redshift and frame dragging \cite{Thorne}.
Throughout, we adopt Eulerian (Zero Angular Momentum Observer (ZAMO)) observers in the Arnowitt–Deser–Misner (ADM) formalism, which provides a natural separation between lapse-driven redshift effects and shift-driven rotational contributions \cite{Arnowitt:1959ah,Gourgoulhon:2012ffd,10.1093/acprof:oso/9780199205677.001.0001}.

A central result of our analysis is that, in the Boyer--Lindquist-based slow-rotation slicing adopted here, the lapse function agrees with its static form up to $\mathcal{O}(a^2)$. Consequently, all redshift effects on Josephson frequencies and DC critical currents remain unchanged relative to the non-rotating case at linear order in $a$. Rotational effects, by contrast, enter through the shift vector and therefore couple to spatial phase transport or to synchronization protocols that refer phase evolution to the global Killing time coordinate. Sagnac-type time delays and their interpretation in relativistic kinematics and stationary spacetimes have been discussed extensively in both conceptual and applied settings~\cite{Anandan:1981zg,Mashhoon:1988zz,Rizzi2004,Ashby2004}. Closely related gravitationally induced quantum phases in interferometry have a long history, beginning with the Colella--Overhauser--Werner neutron experiment and subsequent analyses in general relativity~\cite{Colella:1975dq,Stodolsky:1978ks}. In the present work, this separation has direct physical consequences. The AC Josephson relation retains its redshifted form when expressed in terms of proper voltages, while it reduces to the standard flat-spacetime form when formulated in terms of asymptotic (Killing-time) observables. The DC critical current measured at infinity continues to scale with a single power of the lapse. Interferometric configurations can become sensitive to frame dragging only under additional operational assumptions \cite{Anandan,Ahmedov2005,Jaklevic:1964ysq,Clarke_2004}, for example, when RF-driven phase locking or propagation/synchronization effects relative to the asymptotic time coordinate are relevant. In that case, rotation may induce an effective phase translation of RF interference patterns, while the purely DC fluxoid interference pattern remains unshifted at linear order. By contrast, interferometric configurations such as SQUIDs become sensitive to frame dragging through phase offsets accumulated along different arms of the circuit. The present
work differs in focusing on the gauge-invariant \(3+1\) Josephson phase dynamics
of weak links and SQUID-type interferometry in the slow-rotation regime.

The paper is organized as follows. In Sec. \ref{sec:II} we review the geometry of the slowly rotating spacetimes and their $3+1$ decomposition. Sec. \ref{sec:III} develops the covariant kinematics of superconducting phase dynamics in a stationary background. The AC Josephson effect is analyzed in Sec. \ref{sec:IV}, followed by a discussion of DC transport and critical current scaling in Sec. \ref{sec:V}. In Sec. \ref{sec:VI} we study SQUID interferometry and identify the distinctive signature of frame dragging. We conclude in Sec. \ref{sec:VII} with a summary and outlook toward applications to rotating compact objects and superconducting analogue systems. Throughout this work we use metric signature $(-,+,+,+)$, set $G=c=1$, and keep $\hbar$ explicit.

\section{Geometry of Slowly Rotating Spacetimes: Static Limit and Rotational Corrections}\label{sec:II}

In this section, we briefly review the geometric structure of slowly rotating stationary spacetimes and introduce the corresponding $3+1$ decomposition in the slow-rotation approximation. Our goal is to identify which geometric elements enter the Josephson dynamics at linear order in the rotation parameter. Throughout this work, we restrict our attention to the slow-rotation regime $a/M \ll 1$ and to the exterior region outside the horizon. While static observers cease to exist inside the ergoregion, the $3+1$ decomposition and the Eulerian (ZAMO) congruence remain well defined in the exterior stationary region considered here.

\subsection{Kerr metric and slow-rotation expansion}

The Kerr spacetime describes the exterior gravitational field of a rotating compact object of mass $M$ and angular momentum $J=aM$. In Boyer--Lindquist coordinates $(t,r,\theta,\phi)$, the line element reads \cite{Kerr:1963ud,Visser2007}
\begin{eqnarray}
    \nonumber
ds^2 = -\left(1-\frac{2Mr}{\Sigma}\right)dt^2 &-&\frac{4Mar\sin^2\theta}{\Sigma}\,dt\,d\phi +\frac{\Sigma}{\Delta}dr^2  \\ \label{eq:the_kerr_metric}
&+&\Sigma\,d\theta^2 +\frac{A\sin^2\theta}{\Sigma}\,d\phi^2 ,
\end{eqnarray}

where
\begin{align}
\Sigma = r^2 + a^2\cos^2\theta,
\qquad
\Delta = r^2 - 2Mr + a^2, \nonumber
\\ A = (r^2+a^2)^2 - a^2\Delta\sin^2\theta .
\label{eq:metric_explanation}
\end{align}

{\color{black} In this work, we restrict attention to the slow-rotation regime $a/M \ll 1$ and retain terms only up to linear order in the rotation parameter $a$. At this order, the Kerr geometry is stationary but not static, and the only $\mathcal{O}(a)$ correction appears in the off-diagonal component $g_{t\phi}$, while the diagonal components receive corrections only at $\mathcal{O}(a^2)$.

Expanding~\cref{eq:the_kerr_metric,eq:metric_explanation} to first order in $a$, the metric reduces 
\begin{eqnarray}\nonumber
    ds^{2} &=& -\left(1-\frac{2M}{r}\right)dt^{2}+\left(1-\frac{2M}{r}\right)^{-1}dr^{2}+r^{2}d\theta^{2}  \\ \label{eq:slowrot_metric}
&+&r^{2}\sin^{2}\theta\, d\phi^{2} -\frac{4aM}{r}\sin^{2}\theta \, dt\, d\phi+\mathcal{O}(a^{2}).
\end{eqnarray}

Therefore, to linear order in $a$, the lapse function and the spatial metric coincide with their Schwarzschild forms, while rotational effects enter exclusively through the mixed term $g_{t\phi}$  \cite{Visser2007}.}

We now rewrite the slow-rotation metric \eqref{eq:slowrot_metric} in ADM (3+1) form~\cite{Arnowitt:1959ah,Gourgoulhon:2012ffd,10.1093/acprof:oso/9780199205677.001.0001},
\begin{equation}
ds^{2}=-\alpha^{2}dt^{2}+\gamma_{ij}(dx^{i}+\beta^{i}dt)(dx^{j}+\beta^{j}dt),
\label{eq:ADM}
\end{equation}
where $\alpha$ is the lapse, $\beta^{i}$ is the shift vector, and $\gamma_{ij}$ is the induced spatial metric on $t=\mathrm{const}$ hypersurfaces.

Comparing Eq. \eqref{eq:slowrot_metric} with Eq. \eqref{eq:ADM}, one finds that to linear order in $a$ the lapse and the spatial metric coincide with their Schwarzschild forms,
\begin{equation}
\alpha(r)=\sqrt{1-\frac{2M}{r}}+\mathcal{O}(a^{2}),
\label{eq:lapse_slowrot}
\end{equation}
{\color{black} and 
\begin{equation}
\gamma_{ij}dx^{i}dx^{j}=\left(1-\frac{2M}{r}\right)^{-1}dr^{2}+r^{2}d\theta^{2}+r^{2}\sin^{2}\theta\, d\phi^{2}
+\mathcal{O}(a^{2}).
\label{eq:spatial_metric_slowrot}
\end{equation}
{\color{black}The only $\mathcal{O}(a)$ novelty is the shift. We denote by $\beta^i$ the contravariant shift appearing in Eq. \eqref{eq:ADM}; its covariant form is $\beta_i\equiv\gamma_{ij}\beta^j=g_{ti}$. Therefore,
\begin{equation}
\beta_{\phi}=g_{t\phi}=-\frac{2aM}{r}\sin^{2}\theta+\mathcal{O}(a^{2}),
\label{eq:beta_covariant}
\end{equation}
while the corresponding contravariant component is
\begin{equation}
\beta^{\phi}=\gamma^{\phi\phi}\beta_{\phi}=-\frac{2aM}{r^{3}}+\mathcal{O}(a^{2}),
\label{eq:beta_contravariant}
\end{equation}
since $\gamma^{\phi\phi}=(r^{2}\sin^{2}\theta)^{-1}$ at this order. It is often convenient to introduce the frame-dragging angular velocity
\begin{equation}
\omega_{\rm FD}\equiv -\beta^{\phi}=\frac{2aM}{r^{3}}+\mathcal{O}(a^{2}),
\label{eq:omegaFD}
\end{equation}}
which characterizes the Lense-Thirring dragging of inertial frames in the slow-rotation regime \cite{Lense:1918zz,Mashhoon:1984fj}.

In rotating spacetimes, the notion of ``stationary observers'' is not unique. In particular, static observers following the Killing trajectories $(\partial_t)^\mu$ differ from Eulerian observers that are orthogonal to the $t=\mathrm{const}$ hypersurfaces when the shift vector is nonvanishing ($\beta^i\neq 0$). Since the phase-evolution relations used below depend explicitly on the observer frame, we must fix this choice explicitly.

To make the observer choice explicit, it is convenient to introduce the most general circular four-velocity compatible with stationarity and axisymmetry, written as
\begin{equation}
u^\mu = u^t \Big[(\partial_t)^\mu + \Omega (\partial_\phi)^\mu \Big],
\label{eq:general_circular_velocity}
\end{equation}
where $\Omega \equiv d\phi/dt$ is the angular velocity of the observer with respect to infinity (i.e., with respect to the Boyer--Lindquist time coordinate $t$), and $u^t$ is fixed by the normalization condition $u^\mu u_\mu=-1$.

Static observers correspond to the special case $\Omega=0$; however, this congruence is well-defined only outside the ergoregion, where the Killing vector $(\partial_t)^\mu$ remains timelike. Inside the ergoregion, $(\partial_t)^\mu$ becomes spacelike, and no static observers can exist.

Eulerian observers, on the other hand, are defined as observers orthogonal to the hypersurfaces $t=\mathrm{const}$ \cite{Bardeen:1972fi}. Their angular velocity is given by the frame-dragging rate
\begin{equation}
\Omega_{\rm ZAMO}=\omega_{\rm FD}=-\beta^\phi ,
\label{eq:omega_zamo}
\end{equation}
so that they locally corotate with the dragged inertial frames. More generally, for a corotating superconducting device (e.g., modeling a rotating star or platform), one may set $\Omega=\Omega_\star=\mathrm{const}$.

Throughout this work, we adopt the Eulerian observers, whose four-velocity is the future-pointing unit normal to the hypersurfaces $\Sigma_t$,
\begin{equation}
n^\mu = \frac{1}{\alpha}\left(1,-\beta^i\right),
\qquad
n_\mu = (-\alpha,0,0,0),
\label{eq:zamo_4velocity}
\end{equation}
satisfying $n^\mu n_\mu=-1$. Proper-time intervals measured by these observers are related to the coordinate time by $d\tau = \alpha\,dt$. In the slow-rotation approximation, this choice provides a natural physical interpretation of local measurements, since ZAMO observers locally corotate with the inertial frames dragged by the spacetime.

Note that static observers following the Killing trajectories $(\partial_t)^\mu$ cease to exist inside the ergoregion, whereas ZAMO (Eulerian) observers remain well-defined outside the horizon.

The ADM ($3+1$) decomposition provides a transparent separation between gravitational redshift effects, encoded in the lapse function $\alpha$, and rotational frame-dragging effects, encoded in the shift vector $\beta^{i}$. Since $\alpha(r)=\sqrt{1-2M/r}+\mathcal{O}(a^{2})$, all purely redshift-driven modifications of the Josephson frequency and the asymptotic DC current scaling remain identical to the Schwarzschild case at linear order in $a$.

By contrast, rotation enters the dynamics through the nonvanishing shift component $\beta^\phi$, or equivalently through the frame-dragging angular velocity $\omega_{\rm FD}\equiv -\beta^\phi = 2aM/r^{3}+\mathcal{O}(a^{2})$. Consequently, rotational corrections to Josephson phase evolution arise through the gravitomagnetic coupling term $\beta^{i}p_{i}=\beta^{\phi}p_{\phi}$ whenever the condensate carries a nonvanishing azimuthal momentum component.

This clean separation between lapse-driven redshift and shift-driven frame dragging will play a central role in the analysis of both AC and DC Josephson effects, as well as in SQUID interferometry in the following sections. Throughout this work, we consistently use $\beta^i$ to denote the contravariant shift vector, while $\beta_i=g_{ti}$ denotes its covariant component.
\section{Covariant Josephson kinematics in stationary spacetimes}\label{sec:III}

We now develop the covariant description of superconducting phase dynamics in a stationary spacetime \cite{PantigOvgun2026}. The formulation is based on gauge-invariant phase gradients and conserved currents, and applies equally to static and slowly rotating backgrounds. It provides the kinematic framework used below for the AC and DC Josephson effects.

The superconducting order parameter is written as
\begin{equation}
\Psi = |\Psi| e^{i\theta},
\end{equation}
where $\theta$ is the condensate phase. Throughout this work, $q=2e$ denotes the Cooper-pair charge, with $e>0$ the elementary charge magnitude. The fundamental gauge-invariant quantity governing phase dynamics is the condensate four-momentum \cite{PantigOvgun2026,Aharonov:1959fk}
\begin{equation}
p_\mu = \hbar \nabla_\mu \theta - q A_\mu ,
\label{eq:condensate_momentum}
\end{equation}
where $A_\mu$ is the electromagnetic four-potential.

Let $u^\mu$ denote the four-velocity of the superconducting condensate. In the local rest frame of the condensate, the Josephson relation takes the covariant form
\begin{equation}
u^\mu p_\mu = \mu ,
\label{eq:local_josephson_covariant}
\end{equation}
where $\mu$ is the gauge-invariant electrochemical potential per Cooper pair. With our metric signature $(-,+,+,+)$, this $\mu$ is a convention-defined electrochemical quantity entering the phase-evolution law; it should not be confused with the positive local energy, which is often defined with the opposite sign.

In a stationary spacetime, asymptotic observables are naturally referred to the timelike Killing vector
\begin{equation}
\xi^\mu = (\partial_t)^\mu .
\end{equation}
In the ADM $(3+1)$ decomposition, one writes
\begin{equation}
\xi^\mu = \alpha n^\mu + \beta^\mu ,
\label{eq:xi_adm}
\end{equation}
where $\alpha$ is the lapse, $n^\mu$ is the future-pointing unit normal to the spatial hypersurfaces, and $\beta^\mu$ is the shift vector, with $\beta^\mu n_\mu=0$. In the slow-rotation expansion used here, $\beta^i\beta_i=\mathcal{O}(a^2)$, so that
\begin{equation}
-\xi^\mu\xi_\mu = \alpha^2 - \beta^i\beta_i
\qquad \Rightarrow \qquad
\alpha = \sqrt{-\xi^\mu\xi_\mu} + \mathcal{O}(a^2).
\end{equation}

For superconducting banks at rest with respect to Eulerian observers, we set
\begin{equation}
u^\mu = n^\mu ,
\end{equation}
so that Eq.~\eqref{eq:local_josephson_covariant} becomes
\begin{equation}
\mu = n^\mu p_\mu .
\label{eq:mu_eulerian}
\end{equation}
The invariant temporal component of the condensate momentum is then
\begin{equation}
p_t \equiv \xi^\mu p_\mu
= \hbar\,\partial_t\theta - qA_t .
\label{eq:kv_component}
\end{equation}
Using Eq.~\eqref{eq:xi_adm}, we obtain
\begin{equation}
p_t = \alpha\,\mu + \beta^i p_i ,
\label{eq:phase-evolution-general}
\end{equation}
where
\begin{equation}
p_i = \gamma_i{}^\mu p_\mu
= \hbar \partial_i\theta - qA_i
\label{eq:spatial_momentum}
\end{equation}
is the spatial condensate momentum on the constant-$t$ hypersurfaces.

Equation~\eqref{eq:phase-evolution-general} is the gauge-invariant phase-evolution law in the stationary $(3+1)$ formalism. It shows explicitly that the physically meaningful temporal quantity is not $\partial_t\theta$ by itself, which is gauge dependent, but the combination
\begin{equation}
p_t = \hbar\,\partial_t\theta - qA_t .
\end{equation}
The first term on the right-hand side of Eq.~\eqref{eq:phase-evolution-general} encodes gravitational redshift through the lapse, while the second represents a kinematic contribution associated with the shift vector. In static spacetimes, $\beta^i=0$ and only the lapse term remains \cite{PantigOvgun2026,Aharonov:1959fk}. In rotating spacetimes, however, $\beta^i\neq0$ and an additional rotational contribution appears.

In the slow-rotation Kerr geometry, the only nonvanishing shift component is $\beta^\phi=-\omega_{\rm FD}$, so that
\begin{equation}
\beta^i p_i = \beta^\phi p_\phi .
\end{equation}
This term is nonzero only when the condensate carries a spatial momentum component along the azimuthal direction. Physically, this may occur in the presence of an imposed circulating current, corotation of the device with angular velocity $\Omega_\star$, or persistent loop currents in an interferometric configuration. By contrast, for a short single junction connecting two superconducting banks at rest with respect to Eulerian (ZAMO) observers and without any imposed azimuthal transport current, it is consistent to take
\begin{equation}
p_\phi \simeq 0 .
\end{equation}
In that case, frame dragging does not contribute to the AC or DC Josephson relations at linear order.

As a consistency check, consider the Minkowski limit,
\begin{equation}
\alpha = 1, \qquad \beta^i = 0 .
\end{equation}
Then Eq.~\eqref{eq:phase-evolution-general} reduces to
\begin{equation}
p_t = \mu .
\end{equation}
For an isolated single junction with bias voltage $V$, this yields the standard flat-spacetime AC Josephson relation
\begin{equation}
\hbar \frac{d}{dt}\Delta\phi = 2eV .
\end{equation}

Unless explicitly stated otherwise, all angular frequencies in this work, including the Josephson frequency and the RF drive frequency, are defined with respect to the Killing time coordinate $t$.

Consider a short Josephson junction connecting two nearby superconducting banks
on the same stationary spatial slice. The gauge-invariant phase difference is
defined as
\begin{equation}
\Delta\phi
=
\theta_2-\theta_1-\frac{q}{\hbar}\int_{\mathcal C}A_i\,dx^i ,
\label{eq:DeltaPhi_def_gi}
\end{equation}
where the contour $\mathcal C$ lies on a fixed $t=\mathrm{const}$ hypersurface
and crosses the weak link once.

Differentiating Eq.~\eqref{eq:DeltaPhi_def_gi} with respect to the Killing time
$t$, for a fixed contour $\mathcal C$, gives the exact gauge-invariant identity
\begin{equation}
\hbar\,\frac{d}{dt}\Delta\phi
=
\Delta p_t
-
q\int_{\mathcal C}F_{ti}\,dx^i ,
\label{eq:delta-phi-master}
\end{equation}
where
\begin{equation}
\Delta p_t
\equiv
\left(\hbar\partial_t\theta-qA_t\right)_2
-
\left(\hbar\partial_t\theta-qA_t\right)_1 .
\end{equation}
Using Eq.~\eqref{eq:phase-evolution-general}, this becomes
\begin{equation}
\hbar\,\frac{d}{dt}\Delta\phi
=
\Delta\!\left(\alpha\mu_{\rm ec}+\beta^i p_i\right)
-
q\int_{\mathcal C}F_{ti}\,dx^i .
\label{eq:delta-phi-general}
\end{equation}

Equation~\eqref{eq:delta-phi-general} is the correct starting point for Josephson phase dynamics in a stationary spacetime. It is manifestly gauge invariant. For the short isolated junction considered in this work, we assume that no additional explicitly time-dependent magnetic flux threads the chosen contour and that the local driving is fully encoded in the bankwise electrochemical biases. Writing
\begin{equation}
qV_i^{\rm proper}\equiv \mu_{{\rm ec},i},
\qquad
V_i^\infty\equiv \alpha_i V_i^{\rm proper},
\label{eq:Vi_defs_rot}
\end{equation}
Eq.~\eqref{eq:delta-phi-general} reduces to
\begin{equation}
\hbar\,\frac{d}{dt}\Delta\phi
=
q\left(\alpha_1V_1^{\rm proper}-\alpha_2V_2^{\rm proper}\right)
+\Delta(\beta^ip_i) .
\label{eq:delta-phi-operational}
\end{equation}

In the slow-rotation Kerr geometry, the only nonvanishing shift component is
$\beta^\phi=-\omega_{\rm FD}$, so that
\begin{equation}
\Delta(\beta^i p_i)
=
-\omega_{{\rm FD},1}p_{\phi,1}
+\omega_{{\rm FD},2}p_{\phi,2}.
\label{eq:rotational-term-explicit}
\end{equation}
For a short single junction without azimuthal superflow, $p_{\phi,i}\simeq 0$,
and the rotational contribution vanishes at linear order.

Equations~\eqref{eq:phase-evolution-general} and \eqref{eq:delta-phi-general} make explicit the distinct roles of gravitational redshift and frame dragging. Redshift effects are governed entirely by the lapse function and are identical to those found in static spacetimes at linear order in $a$. Rotational effects arise solely from the shift vector and enter as additional phase terms analogous to a gravitomagnetic or Sagnac contribution.

This separation provides a transparent framework for analyzing both AC and DC Josephson effects in slowly rotating stationary spacetimes. In the following sections, we apply these kinematic relations to derive the redshifted AC Josephson law and examine the influence of frame dragging on interferometric configurations.

\section{AC Josephson effect in slowly rotating stationary spacetimes}\label{sec:IV}

We consider a short Josephson junction whose spatial extent is sufficiently small that the lapse and shift may be treated as constant across the weak link. When discussing the effect of gravitational potential differences, one should interpret the comparison as either comparing two otherwise identical junctions placed at different radii in separate thought experiments or as comparing an interferometric device containing spatially separated junctions. In a strictly local short-junction analysis, the background fields are evaluated at the junction location.

\subsection{Redshifted AC Josephson relation}

We consider a short Josephson junction connecting two superconducting banks located at radii $r_1$ and $r_2$.
The junction is assumed to be at rest with respect to stationary observers and sufficiently small so that the lapse and shift may be treated as constant across its extent.

Starting from the gauge-invariant master equation
\eqref{eq:delta-phi-general}, the time evolution of the Josephson phase
difference is
\begin{equation}
\hbar \frac{d}{dt}\Delta\phi
=
\Delta\!\left(\alpha\mu_{\rm ec}+\beta^i p_i\right)
-
q\int_{\mathcal C}F_{ti}\,dx^i .
\label{eq:ac-general}
\end{equation}
For the short isolated junction considered here, we use the operational
simplification \eqref{eq:delta-phi-operational},
\begin{equation}
\hbar \frac{d}{dt}\Delta\phi
=
q\left(\alpha_1V_1^{\rm proper}-\alpha_2V_2^{\rm proper}\right)
+\Delta(\beta^i p_i) .
\label{eq:ac-with-shift}
\end{equation}
For Cooper pairs, \(q=2e\), so
\begin{equation}
\hbar \frac{d}{dt}\Delta\phi
=
2e\left(\alpha_1V_1^{\rm proper}-\alpha_2V_2^{\rm proper}\right)
+\Delta(\beta^i p_i) .
\label{eq:ac-with-shift-2e}
\end{equation}

When the condensate carries no azimuthal superflow across the single junction,
so that \(p_{\phi,i}\simeq0\), and when no additional explicit Faraday term is
present for the chosen contour, Eq.~\eqref{eq:ac-with-shift-2e} reduces to the
same redshifted AC Josephson relation as in the corresponding static spacetime \cite{Ahmedov2005},
\begin{equation}
\frac{d}{dt}\Delta\phi_{\infty}
=
\frac{2e}{\hbar}
\left(\alpha_1V^{\rm proper}_1-\alpha_2V^{\rm proper}_2\right)
\qquad
\bigl(\mathcal O(a)\bigr).
\label{eq:ac-redshifted}
\end{equation}

{\color{black}Here $\Delta\phi_{\infty}(t)$ denotes the Josephson phase difference parametrized by the Killing time coordinate $t$ (i.e., the time of asymptotic static observers).}

Equation~\eqref{eq:ac-redshifted} shows that, at linear order in $a$, rotation does not modify the gravitational redshift of the Josephson frequency. All redshift effects are fully captured by the lapse function, exactly as in the Schwarzschild case.

\subsection{Rotational contribution to phase evolution}

We now turn to the rotational term $\Delta(\beta^i p_i)$ in Eq. \eqref{eq:ac-with-shift}.
{\color{black} In the slow-rotation approximation, the only nonvanishing contravariant component of the shift vector is $\beta^\phi = -\omega_{\rm FD}$. The rotational contribution, therefore, takes the form
\begin{eqnarray}
    \nonumber
\Delta(\beta^i p_i)& =& \beta^\phi_1 p_{\phi,1} - \beta^\phi_2 p_{\phi,2} \\
&=& -\omega_{{\rm FD},1} p_{\phi,1} + \omega_{{\rm FD},2} p_{\phi,2}.
\end{eqnarray}
}

For a short junction without intrinsic azimuthal superflow, $p_{\phi,i}\simeq 0$, and the frame-dragging contribution vanishes. In this case, the AC Josephson frequency is entirely determined by the redshifted electrochemical potential difference.

More generally, in the presence of a circulating current or in an interferometric configuration, $p_\phi \neq 0$ and rotation induces an additional phase evolution. This contribution is linear in the rotating parameter $a$ and has the structure of a gravitomagnetic or Sagnac-type term \cite{Anandan}. Its effects will be analyzed in detail in Sec. \ref{sec:VI} in the context of SQUID interferometry \cite{AnandanJeeva}.

It is useful to distinguish between two experimentally relevant biasing procedures. If the electrochemical bias is applied locally in the proper frame of each superconducting bank, the Josephson frequency measured at infinity is redshifted according to Eq. \eqref{eq:ac-redshifted}.

Alternatively, if the bias is specified at infinity, the proper voltage satisfies $V_i^{\rm proper} = V_i^{\infty}/\alpha_i$, and Eq. \eqref{eq:ac-redshifted} reduces to the standard flat-spacetime relation,
\begin{equation}
\frac{d}{dt} \Delta\phi_{\infty}=\frac{2e}{\hbar}\left(V_1^{\infty}-V_2^{\infty}\right) .
\end{equation}

Thus, when expressed entirely in terms of asymptotic observables, the AC Josephson relation is insensitive to both gravitational redshift and slow rotation. This invariance underlines the metrological robustness of the Josephson effect, even in a stationary curved spacetime.

The analysis of the AC Josephson effect in slowly rotating spacetimes leads to two central conclusions. First, gravitational redshift \cite{Ummarino2020} enters the Josephson frequency exclusively through the lapse function and is unaffected by rotation at linear order. Second, rotation-specific effects arise only through frame dragging and require nontrivial spatial phase gradients to become observable.

In the absence of such gradients, the AC Josephson effect is insensitive to rotation at linear order and coincides with that of the corresponding static geometry. Rotational signatures, therefore, manifest themselves most clearly in interferometric configurations, which we investigate in subsequent sections.

\section{DC Josephson effect and critical current scaling}\label{sec:V}

We now turn to the DC Josephson effect and examine how the critical current of a weak link is mapped from local (proper) quantities to observables measured at infinity in slowly rotating spacetimes. As in the AC case, our focus is on separating gravitational redshift effects from genuinely rotational contributions.

\subsection{Local current--phase relation}

We consider a short Josephson junction on a stationary spatial slice, connecting two superconducting banks that are comoving with stationary observers. In the tunneling and short-junction limit, the local current--phase relation retains its standard sinusoidal form,
\begin{equation}
I_{\rm proper} = I_c^{\rm proper} \sin\Delta\phi ,
\label{eq:dc-local}
\end{equation}
where $I_c^{\rm proper}$ is the critical current defined in the proper frame of the junction. At this level, the Josephson microphysics is entirely local and independent of the background geometry \cite{Stephenson2023,Ambegaokar:1963zz,Likharev:2022pxz}.

In the slowly rotating spacetimes, the spatial metric on the constant-time hypersurface differs from its Schwarzschild counterpart only at $\mathcal{O}(a^2)$. Consequently, the local current--phase relation and the value of $I_c^{\rm proper}$ are unaffected by rotation at linear order in $a$.

\subsection{Mapping to asymptotic current}

Using the $3+1$ decomposition, the quantity measured by an observer at infinity is the conserved flux of the charge current through the junction per unit Killing time. {\color{black} The current measured by an asymptotic observer is defined as the conserved flux of the charge current through a spatial surface per unit Killing time associated with the stationary Killing vector $\xi^\mu = (\partial_t)^\mu $. The asymptotic current is obtained as the flux of the conserved charge current $j^\mu$ through a two–surface $S$ lying on a $t=\mathrm{const}$ hypersurface,
\begin{equation}
I_\infty = \int_S j^\mu d\Sigma_\mu .
\end{equation}

Since $S$ is a spatial two–surface embedded in $\Sigma_t$, its directed surface element is
\begin{equation}
d\Sigma_\mu = s_\mu \, dS ,
\end{equation}
where $s_\mu$ is the unit spatial normal to $S$ satisfying $s_\mu n^\mu = 0$. The surface area element may be written as
\begin{equation}
dS = \sqrt{\sigma}\, d^2x ,
\end{equation}
where $\sigma$ is the determinant of the induced metric on $S$.

Using the $3+1$ decomposition of the current,
\begin{equation}
j^\mu = \rho n^\mu + J^\mu , \qquad J^\mu n_\mu = 0 ,
\end{equation}
the flux becomes
\begin{equation}
I_\infty = \int_S J^\mu s_\mu \, dS .
\end{equation}

Starting from the covariant conservation law
\begin{equation}
\nabla_\mu j^\mu =0 ,
\end{equation}

its 3+1 decomposition gives
\begin{equation}
\partial_t(\sqrt{\gamma}\rho)
+
\partial_i\!\left[\sqrt{\gamma}(\alpha J^i-\rho\beta^i)\right]
=0 .
\end{equation}

The spatial flux through a surface $S$ with unit normal $s_i$
is therefore
\begin{equation}
I_\infty =
\int_S
\sqrt{\sigma}(\alpha J^i-\rho\beta^i)s_i d^2x .
\end{equation}

In the stationary DC regime relevant for a short junction, charge accumulation is negligible ($\rho \simeq 0$). The flux, therefore, reduces to
\begin{equation}
I_\infty =
\int_S \alpha J^i s_i \sqrt{\sigma}\, d^2x .
\end{equation}

For a short junction, the lapse $\alpha$ is approximately
constant across the junction region, allowing it to be
taken outside the integral,
\begin{equation}
I_\infty = \alpha \int_S J^i s_i \sqrt{\sigma}\, d^2x .
\end{equation}

Defining the locally measured proper current as
\begin{equation}
I_{\rm proper} =\int_S J^i s_i \sqrt{\sigma}\, d^2x ,
\end{equation}
we obtain
\begin{equation}
I_\infty = \alpha I_{\rm proper}.
\label{eq:current-mapping}
\end{equation}
}
Substituting Eq. \eqref{eq:dc-local} into Eq. \eqref{eq:current-mapping}, the DC current measured at infinity
becomes
\begin{equation}
I_\infty = \alpha \, I_c^{\rm proper} \sin\Delta\phi .
\end{equation}

The corresponding critical current, therefore, scales as
\begin{equation}
I_{c,\infty} = \alpha \, I_c^{\rm proper} .
\label{eq:dc-scaling}
\end{equation}

Equation~\eqref{eq:dc-scaling} is identical to the result obtained in a static Schwarzschild spacetime and holds to linear order in the rotation parameter.

{\color{black} In DC measurements, we adopt the following protocol: the local junction parameters--specifically the critical current $ I_c^{\mathrm{proper}} $ in the proper frame of each bank---are fixed by the microphysics of the weak link. The asymptotic current $ I_\infty $ is then measured by an observer at infinity using Killing time $ t $. This protocol mirrors the ``locally imposed bias'' case in the AC Josephson effect: local proper quantities are held fixed, and their redshifted values are observed at infinity.}

It is important to emphasize that rotation does not introduce any additional scaling of the DC critical current at $\mathcal{O}(a)$ within the stationary short-junction setup considered here. The lapse function governing the redshift of time intervals is unchanged at this order, while the shift vector does not contribute to stationary DC transport in the absence of relevant spatial phase gradients or charge accumulation.

Frame-dragging effects enter only through time-dependent phase evolution or through couplings to nontrivial spatial condensate momentum, neither of which is present in a purely DC single-junction configuration under our assumptions. As a result, rotation leaves the DC critical-current scaling unaltered at linear order.

The result~\eqref{eq:dc-scaling} admits a clear operational interpretation. The critical current is a flux measured per unit Killing time, and the factor of $\alpha$ arises from the redshift between proper time at the junction and the time coordinate of an observer at infinity.

By contrast, the local Josephson relation~\eqref{eq:dc-local} involves only proper quantities and is insensitive to both gravitational redshift and rotation. This separation between local constitutive physics and global redshift effects mirrors the behavior observed in the AC Josephson relation and underscores the universality of single-$\alpha$ scaling in DC transport in stationary spacetimes.

In summary, in the slow-rotation regime, the DC Josephson effect exhibits the same redshift scaling as in the corresponding static spacetime. The critical current measured at infinity scales with a single power of the lapse function, while rotational effects are absent at linear order.

As in the AC case, this confirms that rotational signatures arise only in configurations sensitive to phase accumulation along extended paths. Such effects are most naturally probed in interferometric devices, which we analyze in the next section.

\section{SQUID interferometry and frame-dragging effects}\label{sec:VI}

We now consider interferometric configurations in which the influence of rotation becomes physically observable. In contrast to single-junction setups, a SQUID geometry allows for the accumulation of phase differences along spatially extended paths, thus providing direct sensitivity to frame dragging in a rotating spacetime \cite{Jaklevic:1964ysq,Clarke_2004}.

\subsection{Phase dynamics in a SQUID loop}

We consider a dc-SQUID consisting of two Josephson junctions labeled $i=1,2$, located at different radii in slowly rotating spacetimes. Each junction obeys the local current--phase relation 
\begin{equation}
I_i^{\rm proper} = I_c^{\rm proper} \sin\phi_i ,
\end{equation}
where $\phi_i$ is the gauge-invariant phase difference across the $i$-th junction.

Flux quantization around the SQUID loop yields the constraint
\begin{equation}
\phi_1-\phi_2 = 2\pi \frac{\Phi}{\Phi_0} + \delta,
\label{eq:fluxoid}
\end{equation} 
where $\Phi$ is the magnetic flux threading the loop, $\Phi_0=h/2e$ is the flux quantum, and $\delta$ denotes a possible non-gravitational static phase offset (e.g., geometric imbalance or junction asymmetry). 
{\color{black} The fluxoid quantization condition follows from the single-valuedness of the condensate phase. Using the gauge-invariant momentum
\begin{equation}
p_\mu = \hbar \nabla_\mu \theta - qA_\mu ,
\end{equation}
one obtains
\begin{equation}
\oint_C (p_\mu + qA_\mu) dx^\mu = 2\pi\hbar n .
\end{equation}

Since the contour C lies on a spatial hypersurface $\Sigma_t$,
we have $dx^\mu = (0,dx^i)$, and therefore
\begin{equation}
\oint_C (\hbar\partial_i\theta - qA_i) dx^i = 2\pi\hbar n .
\end{equation}

Thus, the DC fluxoid constraint involves only spatial components of the condensate momentum. Frame-dragging effects associated with the shift vector $\beta^i$ enter only through the temporal component $p_t$ and therefore do not modify the DC flux quantization at linear order.}

{\color{black} \subsection{Frame-dragging phase as an operational time delay}

{\color{black} In contrast to the DC fluxoid constraint discussed above, the RF response of a SQUID is governed by the temporal evolution of the Josephson phase. In a stationary spacetime, the phase evolution contains an additional contribution associated with the shift vector $\beta^i$, which represents frame dragging. As a result, interferometric configurations become sensitive to rotation through synchronization effects with respect to the global time coordinate.

To define an operationally meaningful frame-dragging contribution,
we follow the standard Sagnac-type construction in stationary spacetimes. Consider a closed contour C lying in the equatorial plane $\theta=\pi/2$ at fixed radius $r$.}

Along the loop, we set $dr=d\theta=0$ and consider null propagation ($ds^2=0$). The metric then reduces to
\begin{equation}
0 = g_{tt}dt^2 + 2g_{t\phi}dt\,d\phi + g_{\phi\phi}d\phi^2 .
\end{equation}
Solving this quadratic equation for $dt$ yields two branches $dt_\pm$, corresponding to the co-rotating and counter-rotating directions. The leading gravitomagnetic contribution to the time difference is obtained by subtracting the two solutions, giving the standard result.}
{\color{black}In a stationary spacetime, the Sagnac coordinate-time difference for counter-propagating null signals
around a closed contour $C$ is \cite{Anandan:1981zg,Mashhoon:1988zz,Rizzi2004,Ashby2004}
\begin{equation}
\Delta t \equiv t_{\rm co}-t_{\rm counter} \;=\; -2\oint_C \frac{g_{0i}}{g_{00}}\,dx^i .
\label{eq:FD_time_delay}
\end{equation}
The sign of $\Delta t$ flips under reversing the orientation of $C$; in what follows, we use the positive magnitude
$\Delta t_{\rm FD}\equiv|\Delta t|$.

For an equatorial loop ($\theta=\pi/2$) at fixed $r$ in the slow-rotation Kerr metric,
$g_{t\phi}\simeq -2aM/r$ and $g_{tt}\simeq -(1-2M/r)$, hence
\begin{equation}
\Delta t_{\rm FD}\equiv|\Delta t|\simeq \frac{8\pi aM}{r(1-2M/r)}. 
\label{eq:time_delay}
\end{equation}}
{\color{black}Therefore, the corresponding Josephson phase shift becomes
\begin{equation}
\Delta\psi_{\rm FD} = \omega_J \Delta t_{\rm FD},
\label{eq:omega_time_delay}
\end{equation}
which is manifestly dimensionless. {\color{black}In an RF-driven measurement, the relevant phase is set by the external drive.
Denoting the RF drive angular frequency by $\Omega_{\rm rf}$, the corresponding drive-phase offset between the two arms is
$\Omega_{\rm rf}\Delta t_{\rm FD}$. On the $n$-th Shapiro step, phase locking implies $\omega_J=n\Omega_{\rm rf}$,
so Eq. (\ref{eq:omega_time_delay}) may equivalently be written as $\Delta\psi_{\rm FD}=n\Omega_{\rm rf}\Delta t_{\rm FD}$.}

At this point, it is important to distinguish between two conceptually different phase constraints. In the purely DC regime, the SQUID phase relation follows from flux quantization of the condensate phase around a closed superconducting loop on a constant-$t$ hypersurface. This constraint depends only on the electromagnetic flux threading the loop and does not involve any propagation or synchronization effects associated with the coordinate time.

By contrast, under RF irradiation, the observed interference envelope is determined by phase locking to the external drive. In this case, the relevant observable is the phase accumulated with respect to the same asymptotic time coordinate used to define the Josephson frequency. Since frame dragging modifies the relative travel time of signals propagating along the two SQUID arms, it produces an additional effective phase offset $\Delta\psi_{\rm FD}=\omega_J\Delta t_{\rm FD}$, which can translate the RF-driven interference pattern even when the DC lobes remain unshifted.}

{\color{black}
\subsection{Numerical estimates}

\paragraph{Neutron star estimate.}
We estimate the magnitude of the frame-dragging contribution for a compact-object setting using Eq. (\ref{eq:time_delay}). Throughout this estimate, we employ geometric units $G=c=1$, such that the mass parameter $M\equiv GM_{\rm phys}/c^2$ has dimensions of length. For a canonical neutron star with $M_{\rm NS}=1.4\,M_\odot$, we have $M\simeq 2.07~{\rm km}$, and we take the loop radius to be $r=10~{\rm km}$  \cite{Cozma2025}. The rotation is characterized by the dimensionless spin parameter $\chi\equiv a/M$, and adopting $\chi=0.2$ yields the Kerr parameter $a=\chi M\simeq 0.41~{\rm km}$. Substituting these values into Eq. (\ref{eq:time_delay}), we obtain the Sagnac-type frame-dragging time delay
\begin{equation}
\Delta t_{\rm FD}^{\rm(NS)} \simeq \frac{8\pi aM}{r\left(1-\dfrac{2M}{r}\right)} \approx 3.66\ {\rm km}.
\end{equation}
Restoring physical units using $1~{\rm km}=10^3/c \simeq 3.336\times 10^{-6}{\rm s}$, this becomes
\begin{equation}
\Delta t_{\rm FD}^{\rm(NS)} \approx 1.22\times 10^{-5}\ {\rm s}.
\end{equation}
The corresponding phase offset, defined by Eq. (\ref{eq:omega_time_delay}) as $\Delta\psi_{\rm FD}=\omega_J\Delta t_{\rm FD}$ with $\omega_J=2\pi f$, is therefore
\begin{equation}
\Delta\psi_{\rm FD}^{\rm(NS)} = 2\pi f\,\Delta t_{\rm FD}^{\rm(NS)}.
\end{equation}
For a representative frequency $f=10^4~{\rm Hz}$, we find
\begin{equation}
\Delta\psi_{\rm FD}^{\rm(NS)} \approx 0.76\ {\rm rad}.
\end{equation}
Thus, in neutron-star environments, the rotational (frame-dragging) contribution can generate a potentially non-negligible phase offset in interferometric Josephson configurations.

\paragraph{Earth estimate.}
For comparison, we evaluate the same expression for the Earth. In geometric units ($G=c=1$), the Earth mass parameter is $M_\oplus \equiv GM_\oplus/c^2 \simeq 4.43\times 10^{-6}\,{\rm km}$, while the Kerr parameter is $a_\oplus \equiv J_\oplus/(M_\oplus c)\simeq 3.9\times 10^{-3}\,{\rm km}$. Taking the loop radius to be the Earth radius, $r_\oplus\simeq 6.37\times 10^3\,{\rm km}$, we obtain
\begin{equation}
\Delta t_{\rm FD}^{(\oplus)} \simeq \frac{8\pi a_\oplus M_\oplus}{r_\oplus\left(1-\dfrac{2M_\oplus}{r_\oplus}\right)} \approx 6.82\times 10^{-11}\ {\rm km}.
\end{equation}
Restoring physical units using $1~{\rm km}=10^3/c \simeq 3.336\times 10^{-6}~{\rm s}$ yields
\begin{equation}
\Delta t_{\rm FD}^{(\oplus)} \approx 2.27\times 10^{-16}\ {\rm s}.
\end{equation}
The corresponding phase offset is
\begin{equation}
\Delta\psi_{\rm FD}^{(\oplus)} = 2\pi f\,\Delta t_{\rm FD}^{(\oplus)}.
\end{equation}
For $f=10^4~{\rm Hz}$, we find
\begin{equation}
\Delta\psi_{\rm FD}^{(\oplus)} \approx 1.43\times 10^{-11}\ {\rm rad}.
\end{equation}
Therefore, the frame-dragging contribution is entirely negligible for terrestrial-scale Josephson interferometers, in sharp contrast with the compact-object case.

\paragraph{Frequency dependence and comparison.}

\begin{figure}[t]
  \centering
  \includegraphics[width=\columnwidth]{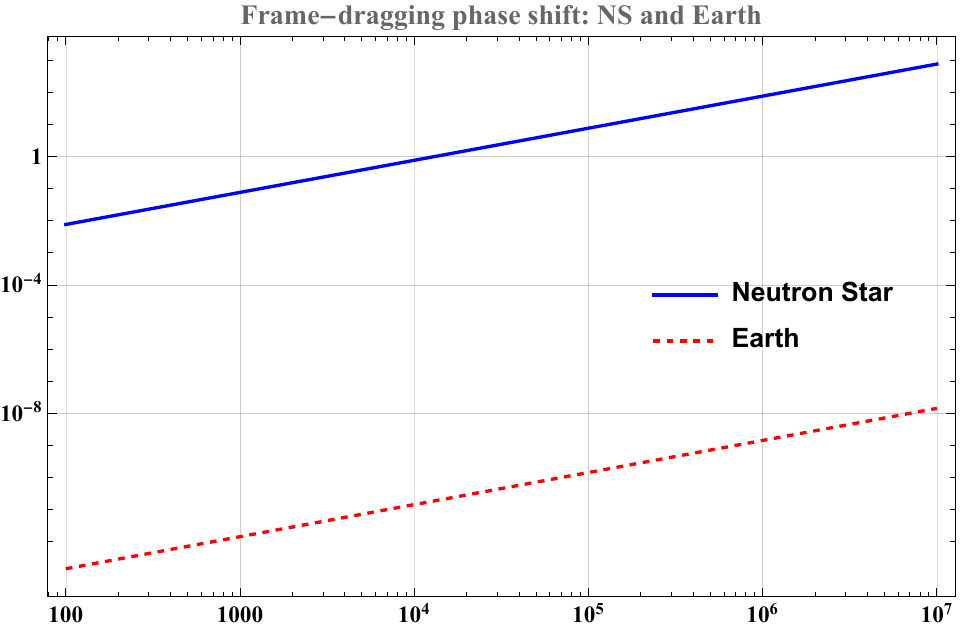}
  \caption{{\color{black}
  Frame-dragging-induced phase shift $\Delta\psi_{\rm FD}$ as a function of the frequency $f$ for an equatorial loop around a neutron star (solid blue curve) and around the Earth (dashed red curve). The phase shift is obtained from the operational relation $\Delta\psi_{\rm FD}=\omega\,\Delta t_{\rm FD}=2\pi f\,\Delta t_{\rm FD}$, where $\Delta t_{\rm FD}$ is the Sagnac-type time delay induced by frame dragging (Eq. \eqref{eq:time_delay}). The neutron-star signal is strongly enhanced by the compactness of the source, whereas the terrestrial phase shift remains negligible over the plotted frequency range.} {\color{black} The legends, tick labels, and axes titles are too small. Please enlarge. You can remove the plot title to be consistent with Fig. 3.}
  \label{fig:NSandEarth}}
\end{figure}

Fig. \ref{fig:NSandEarth} shows the frame-dragging-induced phase shift $\Delta\psi_{\rm FD}$ as a function of the Josephson (or RF-drive) frequency $f$ for both the neutron-star and Earth cases. Using Eqs.~\eqref{eq:time_delay} and \eqref{eq:omega_time_delay}, we plot the phase shift $\Delta\psi_{\rm FD}(f)$ as a function of frequency $f$. The dependence on frequency is strictly linear, $\Delta\psi_{\rm FD}\propto f$, which is consistent with the nearly straight-line behavior in the log-log representation over the range $10^2\le f\le 10^7~{\rm Hz}$.

For neutron-star parameters ($M=1.4M_\odot$, $r=10~{\rm km}$, and $\chi=0.2$ such that $a=\chi M$), the phase shift reaches $\Delta\psi_{\rm FD}\sim 10^{-2}$--$10^{1}~{\rm rad}$ within the plotted frequency window. In contrast, for terrestrial parameters ($M_\oplus\simeq 4.43\times 10^{-6}~{\rm km}$, $a_\oplus\simeq 3.9\times 10^{-3}~{\rm km}$, $r_\oplus\simeq 6.37\times 10^3~{\rm km}$), the phase remains extremely small, $\Delta\psi_{\rm FD}\lesssim 10^{-8}~{\rm rad}$ even at $f=10^7~{\rm Hz}$. This large separation reflects the scaling $\Delta\psi_{\rm FD}\propto (aM/r)\,f$ and demonstrates that frame dragging is irrelevant for Earth-based Josephson interferometers, while it may become non-negligible in compact-object environments. Before proceeding, we stress that the rotational phase shift in the RF response is device-dependent. In particular, its magnitude depends on the geometry, orientation, and effective enclosed area of the interferometer, as well as on the local frame-dragging field. Accordingly, the present analysis is intended as a leading-order covariant framework rather than as a detailed engineering model of a specific SQUID implementation.
}

\subsection{DC SQUID interference pattern}

We first consider the DC regime in the absence of RF driving.
The current measured at infinity is obtained by summing the asymptotic currents through the two junctions,
\begin{equation}
I_\infty = \alpha_1 I_c^{\rm proper} \sin\phi_1 + \alpha_2 I_c^{\rm proper} \sin\phi_2 ,
\end{equation}
where $\alpha_i$ is the lapse evaluated at the location of junction $i$.

Using the constraint~\eqref{eq:fluxoid}, one finds that the SQUID interference pattern depends on the external flux as
{
\begin{equation}
I_\infty(\Phi)= I_c^{\rm proper} \sqrt{\alpha_1^2+\alpha_2^2+2\alpha_1\alpha_2\cos\!\left(2\pi\frac{\Phi}{\Phi_0}+\delta\right)}.
\label{eq:dc_general}
\end{equation}
}
In the symmetric case $\alpha_1=\alpha_2=\bar{\alpha}$, this reduces to
$I_\infty(\Phi)=2\bar{\alpha}I_c^{\rm proper}\left|\cos\left(\pi\Phi/\Phi_0\right)\right|$.

In linear order in the rotation parameter, frame dragging does not generate a constant DC phase offset and therefore does not shift the interference lobes. The leading effect of the gravitational field in the DC regime is instead an overall rescaling of the amplitude through the redshift factor $\bar{\alpha}$.

\subsection{RF-driven SQUID and phase translation}

Under RF irradiation at angular frequency $\Omega_{\rm rf}$, the junction phases acquire an additional time-dependent contribution and phase locking leads to Shapiro steps \cite{Shapiro:1963nhj,Shapiro,Likharev:2022pxz}. For the $n$-th Shapiro step, the effective critical current is modulated by a Bessel factor $J_n(a_{\rm rf})$, where $a_{\rm rf}$ is proportional to the RF drive amplitude. In the presence of a synchronization-induced rotational phase offset $\Delta\psi_{\rm FD}$, the interference envelope can be written phenomenologically as
\begin{eqnarray}
I^{(n)}_{\infty}(\Phi)&=&I_c^{\rm proper}\,|J_n(a_{\rm rf})|\, \\ \nonumber
&\times&
\sqrt{\alpha_1^2+\alpha_2^2+2\alpha_1\alpha_2
\cos\!\left(2\pi\frac{\Phi}{\Phi_0}+\delta+\Delta\psi_{\rm FD}\right)}.
\label{eq:rf_squid_envelope}
\end{eqnarray}
Here $\delta$ denotes a non-gravitational phase offset (e.g., geometric asymmetry or ordinary propagation delay), while $\Delta\psi_{\rm FD}$ represents the additional rotation-induced contribution.

Thus, within this operational RF-synchronization picture, frame dragging enters as an additive phase offset in the RF-driven interference envelope, producing a translation along the flux axis, while the purely DC pattern remains unshifted at linear order.  This behavior is illustrated in Fig. \ref{fig:rf_squid}.

\begin{figure}[t]
  \centering
  \includegraphics[width=\columnwidth]{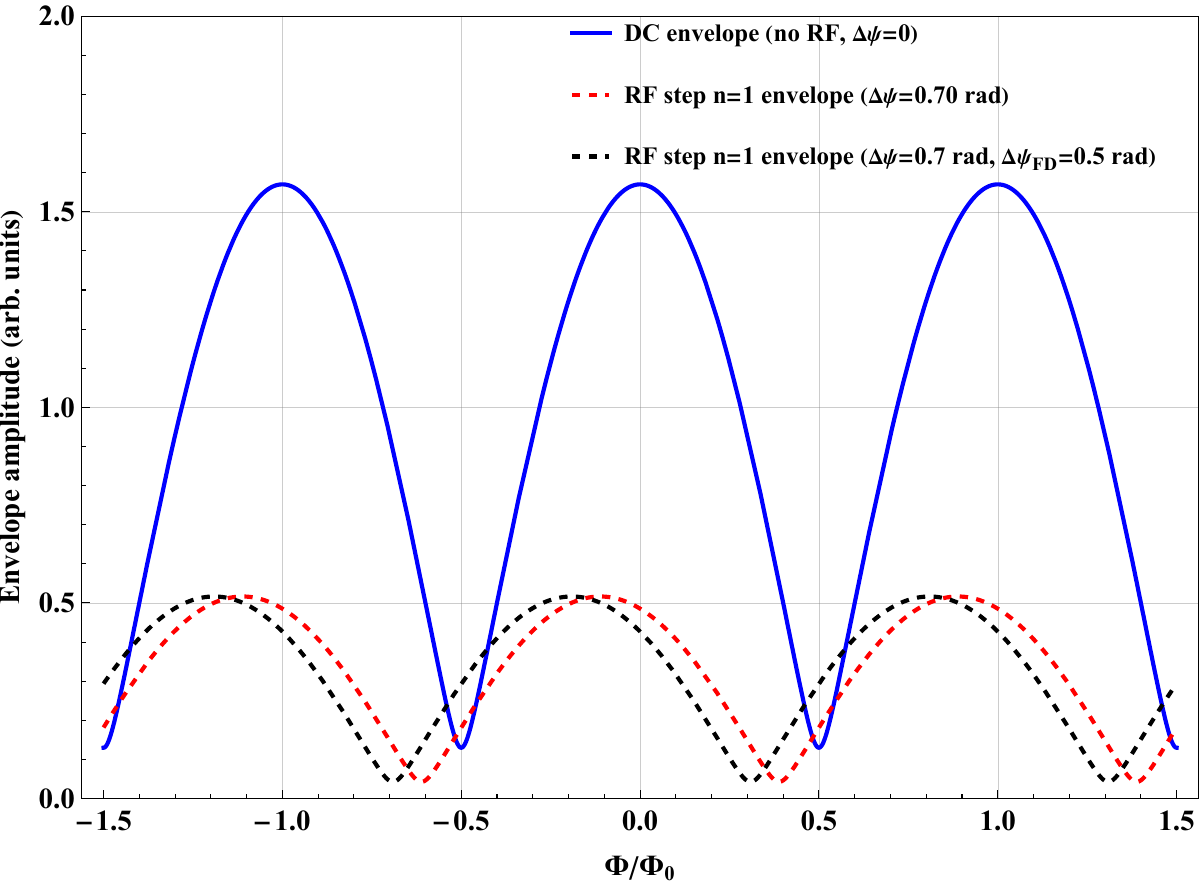}
  \caption{ The legends, tick labels, and axes titles are too small. Please enlarge. Comparison between DC and RF-driven SQUID interference envelopes in the slow-rotation regime. The solid curve shows the DC envelope, which depends only on the magnetic flux $\Phi$ and is not translated by frame dragging at linear order. The dashed curve corresponds to the first Shapiro step ($n=1$) with a finite non-gravitational phase offset $\delta$. The dotted curve includes an additional frame-dragging contribution $\Delta\psi_{\rm FD}=\omega_J\Delta t_{\rm FD}$, producing a further translation of the RF envelope. Gravitational redshift affects only the overall amplitude through the lapse function. In the plot legend, the label $\Delta\psi$ denotes the same non-gravitational offset $\delta$.}
  \label{fig:rf_squid}
\end{figure}

The SQUID analysis highlights a clear separation between redshift and rotational effects. Gravitational redshift, encoded in the lapse function, rescales the overall magnitude of the current measured at infinity but does not modify the purely DC fluxoid relation. By contrast, rotational effects are associated with the shift vector and become observable only in setups sensitive to spatial phase transport or to synchronization with respect to the global Killing time coordinate.

This distinction explains why rotational signatures are absent in local single-junction AC and DC measurements at linear order, while RF-driven interferometric configurations can, in principle, exhibit an additional phase translation.

We now verify that our expressions satisfy the expected limiting cases. In the non-rotating limit $a\to 0$, the frame-dragging angular velocity satisfies $\omega_{\rm FD}\to 0$, and hence any rotation-induced synchronization phase offset vanishes:
\begin{equation}
\Delta\psi_{\rm FD}\to 0 .
\end{equation}
In this limit Eq.~\eqref{eq:dc_general} reduces to
\begin{equation}
I_\infty(\Phi)=I_c^{\rm proper}
\sqrt{\alpha_1^2+\alpha_2^2+2\alpha_1\alpha_2\cos\!\left(2\pi\frac{\Phi}{\Phi_0}+\delta\right)},
\end{equation}
which coincides with the corresponding static-spacetime result.

In the far-field limit $r\to\infty$, one has $\alpha\to1$ and $\omega_{\rm FD}\to0$. For a symmetric SQUID with $\alpha_1=\alpha_2=1$ and vanishing static offset $\delta=0$, Eq.~\eqref{eq:dc_general} becomes
\begin{equation}
I_\infty(\Phi)=2I_c^{\rm proper}\left|\cos\!\left(\pi\frac{\Phi}{\Phi_0}\right)\right|,
\end{equation}
recovering the standard flat-spacetime SQUID response \cite{PantigOvgun2026}.

Finally, the phase offset $\Delta\psi_{\rm FD}$ appearing in Eq.~\eqref{eq:rf_squid_envelope} is dimensionless by construction, since it is defined as the product of an angular frequency and an effective time delay.

\section{Conclusion}\label{sec:VII}

In this work, we have investigated Josephson phenomena in slowly rotating spacetimes, with particular emphasis on disentangling gravitational redshift effects from rotational frame dragging.

Using a covariant formulation based on gauge-invariant phase dynamics and conserved currents in a $3+1$ decomposition, we derived the phase-evolution law appropriate for stationary geometries and applied it to both single-junction and interferometric configurations.

A central result of our analysis is that, to linear order in the rotation parameter $a$, the lapse function coincides with its Schwarzschild form. Consequently, all purely redshift-driven modifications of Josephson frequencies and DC critical currents remain identical to those in the corresponding non-rotating spacetime. The AC Josephson relation retains its standard redshifted structure when expressed in terms of proper voltages and reduces to the usual flat-spacetime form when formulated entirely in asymptotic (Killing-time) observables. Likewise, the DC critical current measured at infinity scales with a single power of the lapse function, $I_{c,\infty}=\alpha I_c^{\rm proper}$, and is unaffected by rotation at $\mathcal{O}(a)$.

Rotational effects arise through the shift vector and therefore require either nontrivial spatial phase gradients or an interferometric protocol sensitive to synchronization with respect to the global time coordinate. For a single junction in the absence of azimuthal superflow, frame dragging does not contribute to either AC or DC measurements at linear order. In a SQUID, the purely DC fluxoid relation remains unshifted at this order, whereas an RF-driven interference pattern can acquire an additional effective phase translation when synchronization or propagation effects are included operationally.

Our results establish a clear separation between lapse-driven redshift effects and shift-driven rotational contributions in Josephson physics. While the weak-field magnitude of frame-dragging effects is expected to be very small for laboratory-scale devices, the formalism developed here provides a consistent framework for analyzing superconducting circuits in stationary spacetimes and offers a starting point for more detailed device-level studies in rotating backgrounds. Extensions to higher-order rotational effects, explicit device geometries, and more general stationary backgrounds constitute natural directions for future work.

\acknowledgments
R. P. and A. \"O. would like to acknowledge networking support of the COST Action CA21106 - COSMIC WISPers in the Dark Universe: Theory, astrophysics, and experiments (CosmicWISPers), the COST Action CA22113 - Fundamental challenges in theoretical physics (THEORY-CHALLENGES), the COST Action CA21136 - Addressing observational tensions in cosmology with systematics and fundamental physics (CosmoVerse), the COST Action CA23130 - Bridging high and low energies in search of quantum gravity (BridgeQG), and the COST Action CA23115 - Relativistic Quantum Information (RQI) funded by COST (European Cooperation in Science and Technology). A. \"O. also thanks to EMU, TUBITAK, ULAKBIM (Turkiye) and SCOAP3 (Switzerland) for their support.

\bibliography{references}

\end{document}